\documentclass{aa}  
\usepackage{graphicx}
\usepackage{txfonts}
\usepackage{natbib}

\bibpunct[]{(}{)}{;}{a}{}{,}
\begin{document}
   \title{The evolutionary state of the southern dense core Cha-MMS1}
   \author{A. Belloche\inst{1}
          \and
          B. Parise\inst{1}
          \and
          F.~F.~S. van der Tak\inst{1}\thanks{Present address: SRON, 9747 AD 
          Groningen, The Netherlands}
          \and
          P. Schilke\inst{1}
          \and
          S. Leurini\inst{1}
          \and
          R. G\"usten\inst{1}
          \and
          L.-{\AA}. Nyman\inst{2}
          }

   \offprints{A. Belloche; belloche@mpifr-bonn.mpg.de}

   \institute{Max-Planck Institut f\"ur Radioastronomie, Auf dem H\"ugel 69,
              DE-53121 Bonn, Germany
          \and
              European Southern Observatory, Alonso de Cordova 3107, Casilla 19001, Santiago 19, Chile
             }

   \date{Received 29 March 2006; accepted  18 May 2006}

 
  \abstract
   {}
   {Our goal is to set constraints on the evolutionary state of the dense core 
  Cha-MMS1 in the Chamaeleon I molecular cloud.}
   {We analyze molecular line observations carried out with the new 
  submillimeter telescope APEX. We look for outflow signatures around the dense
  core and probe its chemical structure, which we compare to predictions of 
  models of gas-phase chemistry. We also use the public database of 
  the Spitzer Space Telescope (SST) to compare Cha-MMS1 with the two Class 0 
  protostars IRAM~04191 and L1521F, which are at the same distance.}
   {We measure a large deuterium fractionation for N$_2$H$^+$ ($11 \pm 3 \%$),
  intermediate between the prestellar core L1544 and the very young 
  Class 0 protostar L1521F. It is larger than for HCO$^+$ 
  ($2.5 \pm 0.9 \%$), which is probably the result of depletion 
  removing HCO$^+$ from the high-density inner region.
  Our CO(3-2) map reveals the presence of a
  bipolar outflow driven by the Class I protostar Ced~110~IRS~4 but we do not 
  find evidence for an outflow powered by Cha-MMS1. We also report the
  detection of Cha-MMS1 at 24, 70 and 160 $\mu$m by the instrument MIPS of the 
  SST, at a level nearly an order of magnitude lower than IRAM~04191 and 
  L1521F.}
   {Cha-MMS1 appears to have already formed a compact object, either 
  the first hydrostatic core at the very end of the prestellar phase, or an 
  extremely young protostar that has not yet powered any outflow, at the very
  beginning of the Class 0 accretion phase.}


   \maketitle
%

\section{Introduction}
\label{s:intro}

The study of the earliest phases of star formation, the so-called prestellar 
and Class 0 stages, is essential to understand the origin of the 
stellar initial mass function \citep*[e.g.][]{Ward-Thompson06}. As one of 
the closest active sites of 
low-mass star formation, the Chamaeleon I dark cloud is an excellent target to 
make progress in this field. However, its population of prestellar cores is 
not known and only 1 or 2 Class 0 protostellar \textit{candidates} have been 
reported so far \citep*[see][]{Reipurth96,Froebrich05}. The most promising of 
them, \object{Cha-MMS1}, is embedded in a C$^{18}$O(1-0) clump 
\citep[][]{Haikala05}, in the Cederblad 110 region where several young stellar 
objects have been identified in the infrared \citep[e.g.][]{Prusti91,Persi01}. 
Using a \textit{tentative} far-infrared detection, \citet{Lehtinen01} derived 
a low temperature (T$_{\mathrm{bol}} = 20$ K) and luminosity 
(L$_{\mathrm{bol}} = 0.45$ L$_\odot$) for Cha-MMS1. They proposed that it is a 
Class 0 protostar, in agreement with \citet*{Reipurth96} who suggested that it 
could be the driving source of the Herbig-Haro objects and the CO outflow seen 
nearby. However \citet{Lehtinen03} failed to detect cm-wave emission with the 
ATCA interferometer, suggesting it is still prestellar.
Here, in an effort to clarify its evolutionary state, we report molecular line 
observations carried out with the 
new submillimeter telescope APEX\footnote{This publication is based on data
acquired with the Atacama Pathfinder Experiment (APEX). APEX is a 
collaboration between the Max-Planck Institut f\"ur Radioastronomie, the 
European Southern Observatory, and the Onsala Space Observatory.} and
the infrared detection of Cha-MMS1 at 24, 70 and 160 $\mu$m by the Spitzer 
Space Telescope.


\section{Observations}
\label{s:obs}
We observed the dense core Cha-MMS1 
\citep*[$\alpha_{2000} = 11^{\mathrm{h}}06^{\mathrm{m}}31\fs7$, 
$\delta_{2000} = -77\degr23\arcmin33\arcsec$,][]{Reipurth96} in July, 
September and November 2005 with the APEX telescope 
(G\"usten et al., \textit{this volume}).
The double-side-band heterodyne receiver APEX-2A 
(Risacher et al., \textit{this volume})
was tuned to the molecular transitions listed in Table~\ref{t:results}.
The N$_2$H$^+$(4-3) line was observed simultaneously with the 
H$_2$D$^+$ line in the 1 GHz bandpass. The half-power beamwidth was 18$\arcsec$
at 345 GHz, and the forward and beam efficiencies used to convert antenna 
temperatures T$_{\mathrm{a}}^\star$ into main-beam temperatures 
T$_{\mathrm{mb}}$ were 0.97 and 0.73, respectively. The 
double-side-band system temperatures 
(continuum calibration) ranged from 100 to 360 K in 
T$_{\mathrm{a}}^\star$ scale. From spectra taken on different days, we estimate
the calibration uncertainty better than $\sim$ 15$\%$. The backend was 
a Fast-Fourier-Transform 
spectrometer with 16384 channels and a channel spacing of 61 kHz yielding an 
effective spectral resolution of about 120 kHz 
(Klein et al., \textit{this volume}).
The telescope pointing was checked 
on  R-Dor, 07454-7112, IRAS~15194-5115, or Mars and found to be accurate to 
$\sim$ 4$\arcsec$ (rms). The telescope focus was optimized on Saturn or Mars. 
The observations were performed in position-switching mode using the APECS 
software 
(Muders et al., \textit{this volume}).
The data were reduced with the CLASS 
software (see http://www.iram.fr/IRAMFR/GILDAS).


\section{Analysis: Kinematics and deuteration}
\label{s:analysis}

\subsection{No evidence for an outflow driven by Cha-MMS1}
\label{ss:outflow}

\begin{figure}[!t]
 \centerline{\resizebox{0.655\hsize}{!}{\includegraphics[angle=270]{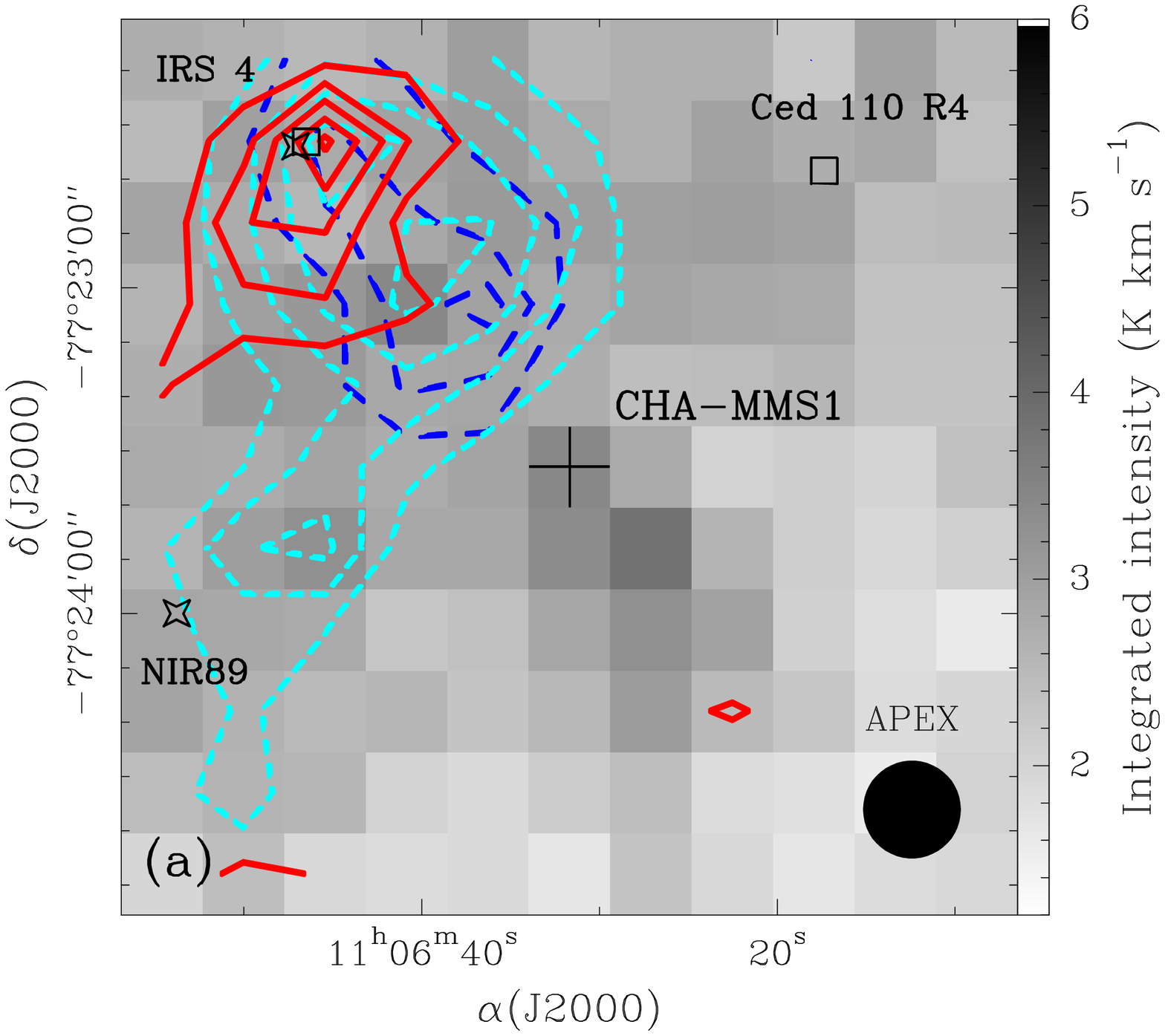}}\hspace*{0.03\hsize}\resizebox{0.315\hsize}{!}{\includegraphics[angle=270]{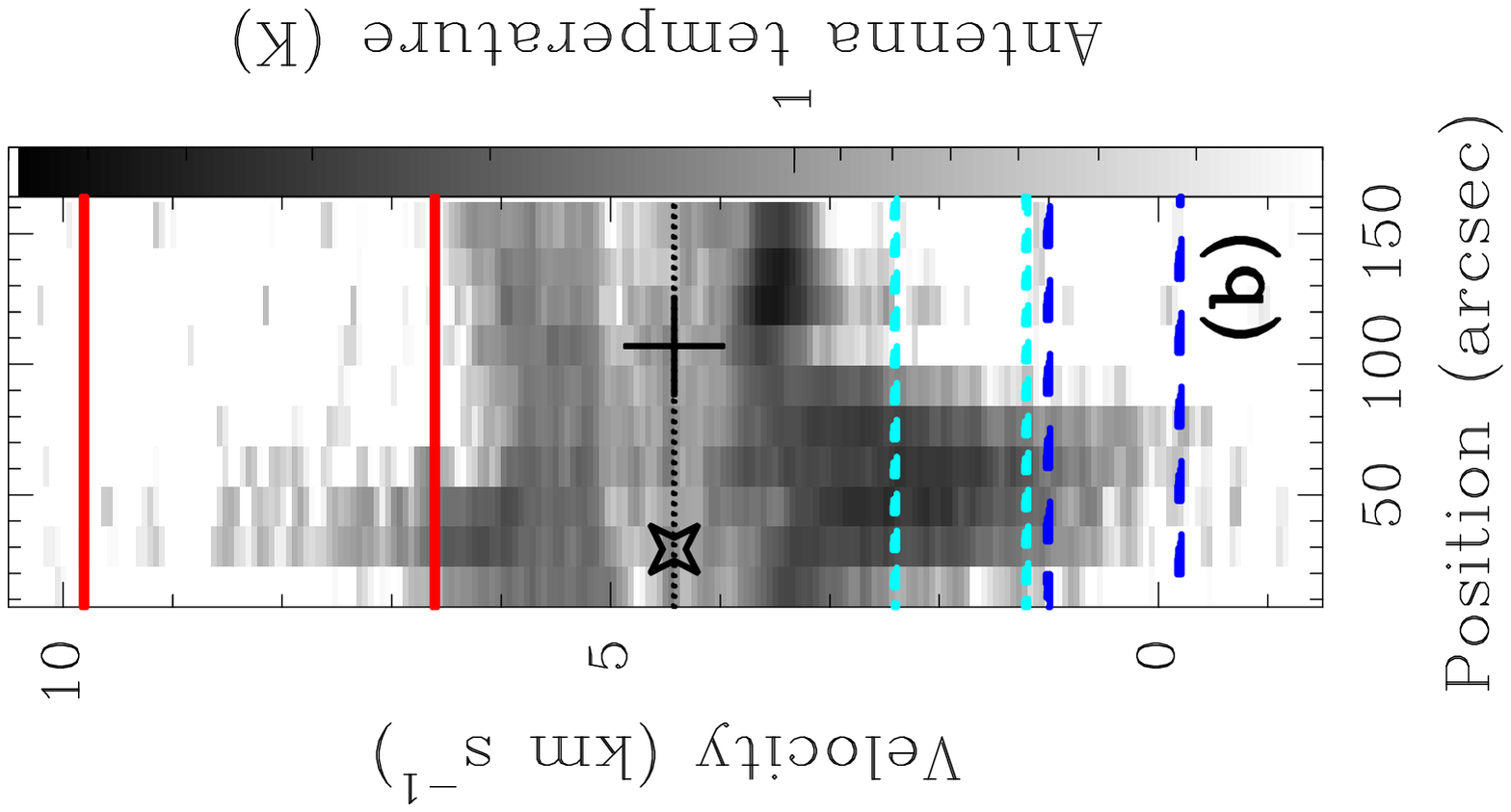}}}
 \vspace*{-1.3ex}
 \caption[]{\textbf{a)} CO(3-2) integrated intensity observed with APEX toward 
 Cha-MMS1 (\textit{cross}). The emission integrated over [3.6,5.8] km~s$^{-1}$
 is shown in greyscale. The blueshifted, less
 blueshifted and redshifted emissions integrated over \hbox{[-0.2,1.0]}, 
 [1.2,2.4] and [6.6,9.8] km~s$^{-1}$ are plotted as long-dashed, short-dashed 
 and thick contours, respectively. The steps are 3$\sigma$, 6$\sigma$ 
 and 3$\sigma$  with rms $\sigma$ = 0.13, 0.13 and 0.24 K~km~s$^{-1}$, 
 respectively. The HPBW is 
 shown on the right. The star and square symbols mark the 
 position of near-infrared \citep[][]{Persi01} and 3.5cm sources
 \citep[][]{Lehtinen03}.
 \textbf{b)}  CO(3-2) position-velocity diagram along the axis going through 
 IRS~4 and Cha-MMS1. The dotted line indicates the systemic velocity of 
 Cha-MMS1. The other lines show the limits of the velocity ranges used for 
 the contour maps. The logarithmic temperature scale varies from 
 0.3 to 6 K.}
 \label{f:co32}
 \vspace*{-1.0ex}
\end{figure}

We mapped the region around Cha-MMS1 in CO(3-2) with a spacing of 15$\arcsec$. 
The spectra are deeply self-absorbed. They are very broad with stronger wing 
emission in
the northeastern part. Figure~\ref{f:co32}a shows maps of the emission 
integrated over the core and the wings of the line. The blueshifted and 
redshifted emissions are clearly 
associated with the Class I protostar \object{Ced~110~IRS~4}
and very likely trace a bipolar outflow. 
Their morphology and the position-velocity diagram shown in 
Fig.~\ref{f:co32}b fall into case 1 of \citet{Cabrit90}, which suggests that 
the outflow axis is close to the line of sight ($i \la 30\degr$). The 
outflow geometry is consistent with the geometry of the bipolar nebula 
seen in the near infrared around IRS~4 \citep[][]{Zinnecker99}. 
The emission integrated
over less blueshifted velocities traces the 
same lobe close to IRS~4 but in addition shows a prominent extension toward
the near-infrared source \object{NIR89}, a potential
Class I young brown dwarf \citep[][]{Persi01}.
This additional wing emission could 
be associated with another outflow driven by NIR89, since \citet{Prusti91} 
detected an elongated redshifted emission in CO(1-0) on the other side of 
NIR89 (see their Fig.~5). This potential second outflow would have an 
intermediate inclination ($20\degr \la i \la 70\degr$) since its two lobes do 
not overlap. On the other hand, we do not find significant wing emission in 
the vicinity of Cha-MMS1, which indicates that it does not drive an 
outflow on the scale probed with APEX.

\subsection{Physical structure of Cha-MMS1}
\label{ss:structure}

We derive the density structure of Cha-MMS1 from 1.3mm continuum 
measurements done with SEST \citep*[][]{Reipurth96}. Assuming a uniform dust 
temperature of 10 K, optically thin emission and a dust opacity 
$\kappa_{\mathrm{1.3mm}} = 0.01$ cm$^{2}$~g$^{-1}$ \citep[][]{vanderTak99},
we derive a mass of 0.40 M$_\odot$ in the 22$\arcsec$ SEST beam and 1.0 
M$_\odot$ in the region within the 5$\sigma$ contour level of 
\citet*{Reipurth96}, which has a deconvolved diameter of $\sim$ 32$\arcsec$. 
If the dense core has a flat inner region and a density 
decreasing as $r^{-2}$, which is typical for prestellar condensations
\citep[e.g.][]{Bacmann00}, then the two measurements yield a central density 
$n_{\mbox{\scriptsize H$_2$,c}} = 3.3^{+0.4}_{-0.6} \times 10^6$ cm$^{-3}$ 
and a radius of the flat inner region $r_{\mathrm{flat}} = 1800^{+450}_{-200}$ 
AU at a distance of 150 pc \citep[][]{Knude98}.

\begin{figure}[!t]
 \centerline{\resizebox{1.0\hsize}{!}{\includegraphics[angle=270]{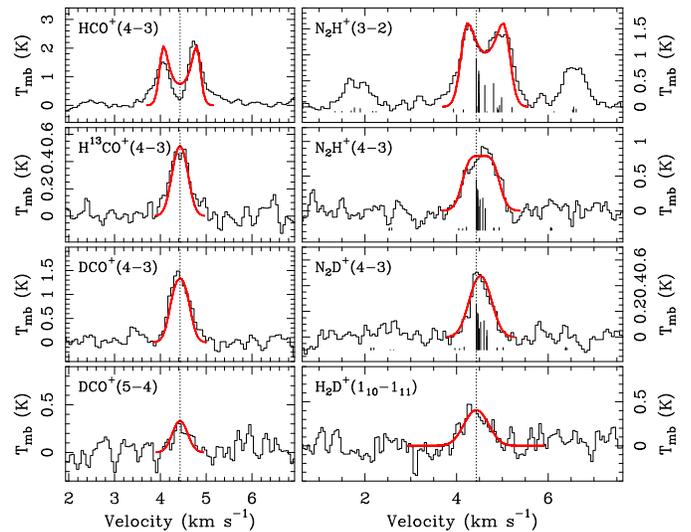}}}
 \vspace*{-1.3ex}
 \caption[]{Spectra observed with APEX toward Cha-MMS1 
 (\textit{histogram}). Synthetic spectra corresponding to the model of 
 Sect.~\ref{ss:dfrac} are superimposed (\textit{thick line}). The 
 hyperfine structure of the N$_2$H$^+$ and N$_2$D$^+$ lines is shown under 
 each spectrum with statistical weighting. 
 The dotted line indicates 
 the systemic velocity assumed for the modeling.}
 \label{f:spectra}
\end{figure}

\subsection{Molecular abundances and deuteration}
\label{ss:dfrac}

The spectra obtained with APEX toward the continuum peak position of Cha-MMS1 
are presented in Fig.~\ref{f:spectra} and the results of single Gaussian and 
7-component HFS fits 
done with the CLASS software are written in Table~\ref{t:fits}. The systemic
velocities derived from the fits agree quite well except for N$_2$H$^+$(4-3)
and H$_2$D$^+$(1$_{10}$-1$_{11}$). Since these two lines were observed 
simultaneously in the same bandpass, we suspect an instrumental problem and 
have shifted these two spectra by -0.20 MHz in Fig.~\ref{f:spectra} to 
compensate for this difference. In the following, we assume a systemic velocity
of 4.43 km~s$^{-1}$. The opacity of N$_2$D$^+$(4-3) is small 
and hence not well 
constrained by the HFS fit. We fixed it to 0.5 (see model below).
The opacity of N$_2$H$^+$(3-2)
is also uncertain, although this is not seen in the statistical 
uncertainty\footnote{\label{fn:abs}The self-absorption of the central group of 
hyperfine components lowers its ``mean peak temperature'' and the opacity 
derived by the HFS fit is then artificially increased. For test purposes, we 
masked out the self-absorption and the opacity was 
reduced to $19.6 \pm 1.1$.}. Assuming a kinetic temperature of 10 K, the 
non-thermal 
linewidths are between $\sim$0.4 and $\sim$0.5 km~s$^{-1}$. The linewidths 
derived from the HFS fits should not be affected by optical depth effects. 
They agree with each other within 2$\sigma$ and we find a weighted mean
$\Delta$V$_{\mathrm{NT}} = 0.40 \pm 0.02$  km~s$^{-1}$. This corresponds to a 
non-thermal dispersion $\sigma_{\mathrm{NT}} = 0.17 \pm 0.01$ km~s$^{-1}$
which is on the same order as the thermal velocity dispersion at 10 K. There 
is therefore a significant amount of non-thermal motions in the dense core.

\begin{table}
\centering
 \caption[]{Results of single Gaussian (\textit{top}) and 7-component
 HFS 
 (\textit{bottom}) fits to the spectra observed toward the center of Cha-MMS1.}
 \label{t:fits}
 \vspace*{-0.0ex}
 \begin{tabular}{lcccccc}
 \hline\hline
 Line & \hspace*{-0.3cm} V$_{\mathrm{lsr}}$$^{(1)}$ & \hspace*{-0.3cm} FWHM$^{(1)}$ & \hspace*{-0.3cm} Area$^{(1)}$ & \hspace*{-0.4cm} T$^\star_{\mathrm{a}}$ & \hspace*{-0.4cm} rms &\hspace*{-0.3cm} $\Delta$V$^{(1,2)}_{\mathrm{NT}}$ \\
  & \hspace*{-0.3cm} {\scriptsize (km/s)} & \hspace*{-0.3cm} {\scriptsize (km/s)} & \hspace*{-0.3cm} {\scriptsize (K~km/s)} & \hspace*{-0.4cm} {\scriptsize (K)} & \hspace*{-0.4cm} {\scriptsize (mK)} & \hspace*{-0.3cm} {\scriptsize (km/s)} \\
 \hline
 H$^{13}$CO$^+$(4-3)          & \hspace*{-0.3cm} 4.42(8) & \hspace*{-0.3cm} 0.53(3) & \hspace*{-0.3cm} 0.20(1)   & \hspace*{-0.4cm} 0.35 & \hspace*{-0.4cm} 41 &  \hspace*{-0.3cm} 0.53(3) \\
 DCO$^+$(4-3)                 & \hspace*{-0.3cm} 4.42(1) & \hspace*{-0.3cm} 0.51(3) & \hspace*{-0.3cm} 0.58(3)   & \hspace*{-0.4cm} 1.08 & \hspace*{-0.4cm} 90 &  \hspace*{-0.3cm} 0.49(3) \\
 DCO$^+$(5-4)                 & \hspace*{-0.3cm} 4.47(4) & \hspace*{-0.3cm} 0.46(8) & \hspace*{-0.3cm} 0.095(15) & \hspace*{-0.4cm} 0.20 & \hspace*{-0.4cm} 71 &  \hspace*{-0.3cm} 0.44(8) \\
 H$_2$D$^+${\scriptsize (1$_{10}$-1$_{11}$)}& \hspace*{-0.3cm} 4.57(4) & \hspace*{-0.3cm} 0.63(7) & \hspace*{-0.3cm} 0.20(2)   & \hspace*{-0.3cm} 0.31 & \hspace*{-0.3cm} 78 & \hspace*{-0.3cm} 0.53(8) \\[0.5ex]
 \hline
 \hline
 Line & \hspace*{-0.3cm} V$_{\mathrm{lsr}}$$^{(1)}$ & \hspace*{-0.3cm} FWHM$^{(1)}$ & \hspace*{-0.3cm} T$^\star_{\mathrm{a}}$$\times \tau_{\mathrm{tot}}$$^{(1)}$ & \hspace*{-0.4cm} $\tau_{\mathrm{tot}}$$^{(1)}$ & \hspace*{-0.4cm} rms & \hspace*{-0.3cm} $\Delta$V$^{(1,2)}_{\mathrm{NT}}$ \\
  & \hspace*{-0.3cm} {\scriptsize (km/s)} & \hspace*{-0.3cm} {\scriptsize (km/s)} & \hspace*{-0.3cm} {\scriptsize (K)} & \hspace*{-0.4cm} & \hspace*{-0.4cm} {\scriptsize (mK)} & \hspace*{-0.3cm} {\scriptsize (km/s)} \\
 \hline
 N$_2$H$^+$(3-2)            & \hspace*{-0.3cm} 4.44(1) & \hspace*{-0.3cm} 0.42(1) & \hspace*{-0.3cm} 24(2) & \hspace*{-0.4cm} 24(2) & \hspace*{-0.4cm} 50 &  \hspace*{-0.3cm} 0.40(1) \\
 N$_2$H$^+$(4-3)            & \hspace*{-0.3cm} 4.65(2) & \hspace*{-0.3cm} 0.50(5) & \hspace*{-0.3cm}  2.3(7)  & \hspace*{-0.4cm} 3.4(14) & \hspace*{-0.4cm} 81 &  \hspace*{-0.3cm} 0.49(5) \\
 N$_2$D$^+$(4-3)            & \hspace*{-0.3cm} 4.44(2) & \hspace*{-0.3cm} 0.41(3) & \hspace*{-0.3cm}  0.53(3) & \hspace*{-0.4cm} 0.5(0)  & \hspace*{-0.4cm} 44 &  \hspace*{-0.3cm} 0.40(3) \\
 \hline
 \end{tabular}
 \vspace*{-1.5ex}
 \begin{list}{}{}
  \item[$^{(1)}$] The statistical uncertainties in parenthesis are given in 
  units of the last digit. 0 means the parameter was fixed. For 
  V$_{\mathrm{lsr}}$, they include the contribution of the frequency 
  uncertainties listed in Table~\ref{t:results}.
  \item[$^{(2)}$] The non-thermal linewidth was computed assuming 
  T$_{\mathrm{K}}$ = 10 K and, for the Gaussian fits, optically thin emission.
 \end{list}
 \vspace*{-3.0ex}
\end{table}

The HCO$^+$(4-3) spectrum is strongly self-absorbed and asymmetric, the red 
peak being stronger than the blue one, and the low-optical depth 
H$^{13}$CO$^+$(4-3) line peaks in between. The asymmetry of the HCO$^+$(4-3) 
line actually changes across our small map: the blue peak is stronger than 
the red one in the southern part below (0,-10$\arcsec$). This pattern 
could result from rotation \citep[e.g.][]{Redman04} or a combination of
infall and rotation, provided the actual 
center of the core is slightly offset from our 
``central'' position toward the South (by $\sim 5\arcsec$). However 
we did not detect a clear overall velocity gradient in our N$_2$H$^+$(4-3) map 
over 50$\arcsec$ that would unambiguously reveal the presence of rotation.

We used the Monte-Carlo-based radiative-transfer code MAPYSO
\citep[see][ and references therein]{Belloche02} to model the
emission in HCO$^+$, H$^{13}$CO$^+$, DCO$^+$, N$_2$H$^+$ and 
N$_2$D$^+$. For all species, we used the collision rates of HCO$^+$ with H$_2$
\citep[][]{Flower99}. We modeled the N$_2$H$^+$ and N$_2$D$^+$ lines as simple
lines, without the hyperfine structure. The H$_2$D$^+$ 
emission was modeled with the Monte-Carlo program RATRAN 
\citep[see][]{Hogerheijde00}. We used the 
density profile described in Sect.~\ref{ss:structure}
and the velocity was set to zero. We assumed uniform gas temperature, abundance
and non-thermal broadening for each molecule. We fixed the isotopic ratio
$\frac{\mbox{\scriptsize [HCO$^+$]}}{\mbox{\scriptsize [H$^{13}$CO$^+$]}}$ to 
70.

The synthetic spectra of the best-fit model are overlaid in 
Fig.~\ref{f:spectra} and the parameters are shown in 
Table~\ref{t:results}. The 
total opacities that we estimate from $\tau_{\mathrm{peak}}$ using the 
statistical weights agree well with the opacities derived from the 
7-component HFS fits (see Table~\ref{t:fits} and 
Footnote~\ref{fn:abs}). The non-thermal 
broadening derived for HCO$^+$ and its isotopologues is more reliable than the 
one \textit{measured} in Table~\ref{t:fits} because the opacity 
broadening is properly taken into account in the modeling. On the other hand, 
it is less reliable for N$_2$H$^+$ and N$_2$D$^+$ since part of it mimics the
broadening by the hyperfine structure not implemented in our model. 
The non-thermal broadening derived for H$_2$D$^+$ is larger than 
for HCO$^+$ and its isotopologues, but given the low signal-to-noise ratio
of the H$_2$D$^+$  spectrum, it may not be significant. The one
\textit{measured} for N$_2$H$^+$ and N$_2$D$^+$ in Table~\ref{t:fits} lies in 
between. From the abundances obtained in Table~\ref{t:results} we derive the 
following deuterium fractionations:
$\frac{\mbox{\scriptsize [N$_2$D$^+$]}}{\mbox{\scriptsize [N$_2$H$^+$]}} 
= 11 \pm 3 \%$ and 
$\frac{\mbox{\scriptsize [DCO$^+$]}}{\mbox{\scriptsize [HCO$^+$]}} = 2.5 \pm 
0.9 \%$.
The deuterium fractionation is therefore $\sim$ 4 times larger for N$_2$H$^+$ 
than for HCO$^+$.

\begin{table}
\centering
 \caption[]{Gas temperature, non-thermal broadening and abundance of each 
 molecule for the best-fit model. The peak opacity of each modeled transition 
 is also given, as well as an estimation of the total opacity of the hyperfine
 multiplets if they had been modeled.}
 \label{t:results}
 \vspace*{-0.0ex}
 \begin{tabular}{lclcccccc}
 \hline\hline
 Molec. & \hspace*{-0.40cm} Line & \multicolumn{1}{c}{\hspace*{-0.32cm} Frequency$^{(1)}$} & \hspace*{-0.32cm} T$_\mathrm{k}$ & \hspace*{-0.32cm} $\Delta$V$_{\mathrm{NT}}$ & \hspace*{-0.32cm} $\chi^{(2)}$ & \hspace*{-0.32cm} $\tau_{\mathrm{peak}}$ & \hspace*{-0.32cm} $\tau_{\mathrm{tot}}$ \\
  & \hspace*{-0.40cm} & \multicolumn{1}{c}{\hspace*{-0.32cm} {\scriptsize (MHz)}} & \hspace*{-0.32cm} {\scriptsize (K)} & \hspace*{-0.32cm} {\scriptsize (km/s)} & \hspace*{-0.32cm} & \hspace*{-0.32cm}  & \hspace*{-0.32cm} \\
 \hline
 HCO$^+$    & \hspace*{-0.40cm} 4-3 & \hspace*{-0.32cm} 356734.134(50) & \hspace*{-0.32cm} 8.8 & \hspace*{-0.32cm} 0.35 & \hspace*{-0.32cm} 8.4(-10)  & \hspace*{-0.32cm} 44 & \hspace*{-0.32cm} - \\
 H$^{13}$CO$^+$ & \hspace*{-0.40cm} 4-3 & \hspace*{-0.32cm} 346998.338(89) & \hspace*{-0.32cm} 8.8 & \hspace*{-0.32cm} 0.35 & \hspace*{-0.32cm} 1.2(-11) & \hspace*{-0.32cm} 0.46 & \hspace*{-0.32cm} - \\
 DCO$^+$    & \hspace*{-0.40cm} 4-3 & \hspace*{-0.32cm} 288143.858(7) & \hspace*{-0.32cm} 8.8 & \hspace*{-0.32cm} 0.35 & \hspace*{-0.32cm} 2.1(-11) & \hspace*{-0.32cm} 1.2 & \hspace*{-0.32cm} - \\
            & \hspace*{-0.40cm} 5-4 & \hspace*{-0.32cm} 360169.778(7) & \hspace*{-0.32cm}   &  \hspace*{-0.32cm}     & \hspace*{-0.32cm}          & \hspace*{-0.32cm} 0.25 & \hspace*{-0.32cm} - \\
 N$_2$H$^+$ & \hspace*{-0.40cm} 3-2 & \hspace*{-0.32cm} 279511.858(11) & \hspace*{-0.32cm} 7.3 & \hspace*{-0.32cm} 0.50 & \hspace*{-0.32cm} 2.2(-10) & \hspace*{-0.32cm} 15 & \hspace*{-0.32cm} 16 \\
            & \hspace*{-0.40cm} 4-3 & \hspace*{-0.32cm} 372672.560(13) & \hspace*{-0.32cm}   & \hspace*{-0.32cm}      & \hspace*{-0.32cm}          & \hspace*{-0.32cm} 3.4 & \hspace*{-0.32cm} 3.6 \\
 N$_2$D$^+$ & \hspace*{-0.40cm} 4-3 & \hspace*{-0.32cm} 308422.322(10) & \hspace*{-0.32cm} 7.3 & \hspace*{-0.32cm} 0.50 & \hspace*{-0.32cm} 2.5(-11) & \hspace*{-0.32cm} 0.49 & \hspace*{-0.32cm} 0.51 \\
 o-H$_2$D$^+$ & \hspace*{-0.40cm} {\scriptsize 1$_{10}$-1$_{11}$} & \hspace*{-0.32cm} 372421.385(10) & \hspace*{-0.32cm} 8 & \hspace*{-0.32cm} 0.50 & \hspace*{-0.32cm} 4.3(-11) & \hspace*{-0.32cm} 0.19 & \hspace*{-0.32cm} - \\
 \hline
 \end{tabular}
 \vspace*{-0.7ex}
 \begin{list}{}{}
  \item[$^{(1)}$] from the CDMS catalog as of Feb. 2006 
    \citep[see][]{Mueller05}. The uncertainties in parenthesis are 
    given in units of the last digit.
  \item[$^{(2)}$] $X$($p$) means $X \times 10^{p}$.
 \end{list}
 \vspace*{0.0ex}
\end{table}


\section{Implications: Evolutionary state of Cha-MMS1}
\label{s:implications}

The deuterium fractionations of HCO$^+$ and N$_2$H$^+$ derived in 
Sect.~\ref{ss:dfrac} are 3-4 orders of magnitude larger than the cosmic
ratio $\frac{\mbox{\scriptsize [D]}}{\mbox{\scriptsize [H]}} \sim 10^{-5}$.
This enhanced molecular deuteration is typical for low-mass dense cores 
\citep[e.g.][]{Williams98,Crapsi05,Parise06}
and correlates well with CO depletion in prestellar cores
\citep[][]{Bacmann03}. It is understood as a result of CO freeze out onto the 
grain surface, CO being the major destroyer of H$_2$D$^+$, the key 
ion in the molecular deuterium chemistry \citep[e.g.][]{Caselli03}. 
To understand the different deuteration degrees measured for 
HCO$^+$ and N$_2$H$^+$ in Cha-MMS1, we compare our results
with the predictions of chemical models. \citet{Roueff05} calculated the 
steady state of models of gas-phase chemistry at 10 K. Their three 
models, Model 1, 2, and 3, have densities of $10^4$, $10^5$, and 
$10^6$ cm$^{-3}$, and C and O depletion factors of 1, 5, and 15, respectively.
 At 10 K, the ortho-to-para ratio of 
H$_2$D$^+$ is close to 1 \citep*[][]{Gerlich02}, so we deduce a total 
H$_2$D$^+$ abundance of $\sim 9 \times 10^{-11}$, in agreement with Model 3 
and close to the value derived for the evolved prestellar core \object{L1544} 
\citep[][]{vanderTak06}.
For N$_2$H$^+$, both the abundance and the deuterium fractionation
in Table~\ref{t:results} agree well with the predictions of Model 3,
while they are close to Model 2 for HCO$^+$. The degree of deuterium
fractionation in HCO$^+$ traces therefore 
lower densities than the N$_2$H$^+$ one. This can be understood if 
HCO$^+$ is depleted at high densities, which is expected in dense cores since 
HCO$^+$ is chemically related to CO. To test this scenario, we computed a 
model for HCO$^+$ and isotopologues with a hole of radius 1800 AU and 
abundances 2.5 times larger: the fit to the observed spectra was as good 
as for the model shown in Fig.~\ref{f:spectra}, suggesting that HCO$^+$ may 
indeed be depleted in the inner region. \citet{Crapsi05} derived a N$_2$H$^+$
deuterium fractionation of 16-23$\%$ for L1544 and 5-10$\%$ for 
\object{L1521F}, which is now known as a Class 0 protostar since its detection 
by the 
Spitzer Space Telescope (SST) in the near infrared \citep[e.g.][]{Terebey05}. 
Our value for Cha-MMS1 lies in between, which suggests an evolutionary state 
between L1544 and L1521F if the 
$\frac{\mbox{\scriptsize [N$_2$D$^+$]}}{\mbox{\scriptsize [N$_2$H$^+$]}}$ 
ratio can be used as a chemical clock.

The higher spatial resolution provided by APEX and the higher densities traced 
by CO(3-2) allowed us to resolve the blueshifted wing emission seen 
earlier in CO(1-0) \citep[][]{Mattila89,Prusti91}
into two outflows -- the second one being 
tentative --, and assign their driving sources. The main outflow is
\textit{not} powered by Cha-MMS1, 
as \citet*{Reipurth96} suggested, but by the Class I protostar Ced~110~IRS~4.
This is consistent with the non-detection of Cha-MMS1 in centimeter continuum 
emission \citep[][]{Lehtinen03}.
Since we did not find evidence for an outflow driven by Cha-MMS1 itself, the 
dense core must be in an earlier evolutionary state than the typical 
young Class 0 protostar \object{IRAM~04191} which does have a 
powerful outflow \citep[][]{Andre99}.
However, using the SST archive (http://ssc.spitzer.caltech.edu/), we 
discovered a source detected at 24, 70 and 160 $\mu$m by the instrument MIPS 
(PID:37, 
REQID:3962112), offset from Cha-MMS1 by 5-7$\arcsec$ toward the East. Given 
the large beams of SEST (22$\arcsec$) and SST (6-18$\arcsec$) and their 
pointing rms accuracies (3$\arcsec$ and 1.4$\arcsec$, respectively), we 
find it conceivable that this mid-infrared source is embedded in the Cha-MMS1
dense core. We measure fluxes of $2.5 \pm 0.8$ and 
$200 \pm 100$ mJy at 24 and 70 $\mu$m. The 24 $\mu$m flux is 
$\sim$ 7 and 10 times smaller than the fluxes we measured for 
IRAM~04191 and L1521F using the SST data products of the C2D legacy program
\citep[][]{Evans03}, 
and the 70 $\mu$m flux $\sim$ 4 and 3 times smaller\footnote{Note that 
the 70 $\mu$m flux is one order of magnitude weaker than the questionable 
3.7$\sigma$ detection of \citet{Lehtinen01} with ISOPHOT.}. In addition, 
Cha-MMS1
was not detected at 8~$\mu$m by the instrument IRAC (SST public data, PID:37,
REQID:3960320), at a 3$\sigma$ level $\sim$ 9 and 7 times lower than the peak 
fluxes of IRAM~04191 and L1521F. The detection at 24~$\mu$m implies that 
Cha-MMS1 already contains a compact hydrostatic object 
\citep[e.g.][]{Masunaga00}, 
but the weakness and ``redder'' color of the 24 and 70~$\mu$m fluxes and the 
non-detection at 8~$\mu$m suggest that it is less evolved than IRAM~04191 and 
L1521F \citep[e.g.][]{Young05}, since they are at the same distance. Depending 
on its inclination along the line of
sight \citep[see][, for its effects on the mid-infrared 
fluxes]{Boss95,Whitney03}, it could be at the stage of the first hydrostatic 
core, i.e. at the very end of the prestellar phase, or have already 
formed an extremely young Class 0 protostar that has not yet powered any
outflow. At a stage intermediate between L1544 and L1521F, Cha-MMS1 would then
be the first object found so close to the very beginning of the Class 0 
accretion phase.

\vspace*{-0.5ex}
\begin{acknowledgements}
We thank P. Andr\'e and J. Kauffmann for enlightening discussions 
about Cha-MMS1 and the SST, and the APEX staff for their 
help during the observations. BP is grateful to the A. von Humboldt Foundation 
for a Humboldt Research Fellowship.
\end{acknowledgements}


\end{document}